\documentclass[twocolumn,letterpaper]{revtex4}

\usepackage{graphicx}
\usepackage{bm}
\usepackage{epsfig}

\begin{document}

\special{portrait}


\title{Truncation of periodic image interactions for confined systems}

\author{Sohrab Ismail-Beigi} \affiliation{Department of Applied Physics,
  Yale University, New Haven, CT 06520}
\date{\today}

\begin{abstract}
First principles methods based on periodic boundary conditions  are used extensively by materials theorists.  However, applying these methods to systems with confined electronic states entails the use of large unit cells in order to avoid artificial  image interactions.  We present a general approach for truncating the Coulomb interaction that removes image effects directly and leads to well converged results for modest-sized periodic cells.  As an illustration, we find the lowest-energy quasiparticle and exciton states in two-dimensional hexagonal GaN sheets.  These sheets have been proposed as parent materials for single-walled GaN nanotubes which may be of interest for optoelectronics.
\end{abstract}


\maketitle

First principles methods, especially those based on Density Functional Theory (DFT) \cite{HK,KS}, are used widely to calculate physical and chemical properties.  One very popular approach uses plane wave (Fourier) basis sets, periodic boundary conditions, and pseudopotentials \cite{RMP}.  Plane waves provide a general, orthonormal basis with controllable convergence.  They also are natural for physical situations where electronic states extend in all directions (e.g.  bulk materials).  Mature, well optimized and parallelized plane-wave software packages are publicly available \cite{somecodes}.  In addition to electronic ground-state properties from DFT, quasiparticle and optical excitations can be calculated with methods that build on this theoretical and computational edifice \cite{TDDFT1,TDDFT2,HL,BSE1,BSE2,BSE3}.

Many systems of interest have spatially confined electronic states.  Molecules are classic cases, but nanostructures such as nanotubes, nanowires, or sheet-like systems are of current interest.  Often, it is the confinement of electronic states that leads to an unusual behavior.  Unfortunately, studying these structures with plane waves requires the use of large unit cells with regions of empty space surrounding the structures.  This is necessary to ensure that periodic copies do not interact: one usually aims to study the isolated structure and not the artificial array.  The electronic wave functions are not a problem because they decay exponentially into the vacuum regions and converge rapidly with increasing periodic separation.  Rather, the main difficulty stems from the long-range Coulomb interaction.  It can lead to long-range monopolar or dipolar image interactions that fall off slowly (algebraically) with periodic separation \cite{MakovPayne,ORGDA}.  In practice, one has to model the image interactions and extrapolate to infinite separation \cite{MakovPayne}.

Ideally, one should eliminate Coulomb interactions among periodic copies from the start.   Truncation of the Coulomb interaction was used for molecular systems and carbon nanotubes \cite{ORGDA,CNTvctrunc}.  Here we provide a general scheme for truncating the Coulomb interaction in systems having both periodic and confined spatial directions.  The scheme should interest researchers investigating confined systems with plane wave methods or, more generally, periodic boundary conditions because it leads to rapid convergence versus periodic separation.  Our illustrative calculations focus on calculating electronic excitations where image effects are most pronounced \cite{ORGDA,RL}.  However, the approach is equally useful and easy to implement for ground-state methods.

We begin with ground-state calculations using plane waves.  The component of the total energy incorporating long-range Coulomb interactions is the Hartree energy,
\begin{eqnarray*}
E_H & = & \frac{1}{2}\int d^3r \int d^3r' \ \rho(r)v_c(r-r')\rho(r')\nonumber\\
 & = & \frac{1}{2}\sum_{G\ne0}|\hat \rho(G)|^2 \hat{v}_c(G)\,,
\end{eqnarray*}
where $v_c(r)$ is the Coulomb kernel $v_c(r)=1/|r|$ for the untruncated case, and $\rho(r)$ is the total charge density.  The $G$ are the reciprocal lattice vectors of the periodic cell, and $\hat\rho(G)$ is the Fourier transform of $\rho(r)$.  The Fourier transform of the untruncated interaction is $\hat v_c(k)=4\pi/k^2$.  (The $G=0$ term is excluded due to a compensating background \cite{RMP}.)  The remaining terms in total energy, the kinetic and exchange-correlation energies, are short-ranged for the standard semi-local approximations such as the LSDA or GGA \cite{RMP}.  We want to replace $v_c(r)$ by a function that explicitly zeroes interactions between periodic images.  Obviously, the replacement must yield a Fourier transform $\hat v_c(k)$ that is well-defined for $k\ne0$.  

For an isolated atom or molecule, we truncate in all three spatial directions.  A simple spherical truncation with radius $r_c$ was used previously \cite{ORGDA,RL},
\begin{equation}
v^{sp}_c(r) = \frac{\theta(r_c-|r|)}{|r|} \rightarrow 
\hat v^{sp}_c(k) = \frac{4\pi}{k^2}\left\{1-\cos(|k|r_c)\right\}
\end{equation}
The interaction $\hat v^{sp}_c(k)$ is finite for any $k$.  The issue is to choose $r_c$.  One takes $r_c$ to be larger than the ``size'' of the molecule $d$ (the region where $\rho(r)$ is appreciable).  Thus intramolecular interactions are calculated correctly.  Simultaneously, $r_c$ must be smaller than the shortest distance $L$ between periodic images minus $d$ to prevent periodic image interactions, leading to $d<r_c<L-d$ \cite{ORGDA,RL}.  A practical solution is to set $r_c=L/2$ and increase $L$ to achieve convergence.

The smoothness of $\hat v_c(k)$ is crucial for calculations of electronic excitations.  Green's function-based methods describing quasiparticle and excitonic dynamics require the screened Coulomb interaction matrix $W_{GG'}(q)$ where $q$ is a wave vector in the first Brillouin zone \cite{HL,RL}.  (We will truncate interactions in the confined directions so $q$ only has components along the non-confined, i.e. physically periodic, directions.) $W$ is calculated through the dielectric matrix $\epsilon$,
\begin{eqnarray*}
W_{GG'}(q) & = & \epsilon^{-1}_{GG'}(q)\,\hat v_c(q+G') \\
\epsilon_{GG'}(q) & = & \delta_{GG'} - \hat v_c(q+G)\,\chi_{GG'}(q)\,,
\end{eqnarray*}
where $\chi$ is the irreducible polarizability typically evaluated within the random-phase approximation (RPA) \cite{HL}.  (The frequency dependences of $W$, $\epsilon$, and $\chi$ are suppressed for clarity.) Calculations of excited state properties require sums over the entries of $W$ and an integral over $q$.  With spherical truncation, $\hat v^{sp}_c(k)$ is well behaved making for straightforward calculation of $W$.  For the usual untruncated case, $\hat v_c(k)=4\pi/k^2$ diverges as $k\to0$, but the key divergence in $W$ is confined to the ``head'' element $G=G'=0$ when $q\to0$ \cite{CPM,HL,insulcomment}. The calculation and inversion of $\epsilon$ is handled separately for this special entry \cite{HL}.  In brief, divergent behavior in $W$ is confined to one known entry that is treated ``by hand''.

We derive analogous truncations for the two remaining cases:  (1) sheet-like geometries with one confined and two periodic directions (e.g. graphene or thin films), and (2) wire-like geometries with one periodic and two confined directions (e.g. nanotubes or nanowires).

We begin with the sheet  geometry.  The confined direction is along the $z$ axis, and the system is periodic in the $xy$ plane.
We choose a truncation length $z_c$ and set
\begin{equation}
v^{sh}_c(r) =\frac{\theta(z_c-|z|)}{|r|}\,.
\end{equation}
We require $d<z_c<L_z$ where $d$ is the the extension of electronic states into the inter-sheet vacuum region and $L_z$ is the periodicity along $z$.  The Fourier transform is
\begin{eqnarray*}
\hat{v}^{sh}_c(k) = \frac{4\pi}{k^2}
\left\{1+e^{-k_{xy}z_c}\left[\frac{k_z}{k_{xy}}\sin(k_zz_c)-\cos(k_zz_c)\right]
\right\}
\end{eqnarray*}
where $k_{xy}=(k_x^2+k_y^2)^{1/2}$.  As a check, in the short-wavelength limit $k_{xy}z_c\gg1$, we recover the untruncated interaction.  
For arbitrary $z_c$, this formula diverges when $k_{xy}\rightarrow0$ for \emph{any} nonzero $k_z$ and is clearly unsuitable.

However, the choice $z_c=L_z/2$ is special.  For the sheet geometry, reciprocal lattice vectors have $k_z=2\pi n_z/L_z$ for integer $n_z$.  Thus $\sin(k_zz_c)=0$ for \emph{all} $n_z$ when $z_c=L_z/2$.
We arrive at the simplification
\begin{eqnarray}
\hat{v}^{sh}_c(k) = \frac{4\pi}{k^2}
\left\{1-e^{-k_{xy}z_c}\cos(k_zz_c)\right\}\,.
\end{eqnarray}
The Coulomb interaction is now finite for any $k_{xy}$ when $k_z\ne0$.  For $k_z=0$, it diverges as $4\pi z_c/k_{xy}$ as $k_{xy}\rightarrow0$.
As expected, the divergence is milder than the untruncated 3D case.  Our choice of $z_c$ has confined the divergence to the single wave vector $k=0$.  This simple replacement for $\hat{v}_c(k)$ allows for straightforward ground-state calculations. (The minor required changes for excited states are outlined below.)

We proceed to the wire geometry.  The periodic direction (wire axis) is along $z$, and the system is confined in the $xy$ plane. We seek a truncation of the form
\begin{equation}
v^{wi}_c(r) = \frac{\theta(x,y)}{|r|}\,.
\end{equation}
The shorthand $\theta(x,y)$ returns zero unless $x$ and $y$ lie in a finite region surrounding the wire that is specified below.  The Fourier transform along $z$ gives
\begin{equation}
\hat{v}^{wi}_c(k) = \int\!\! dx\!\! \int \!\! dy\, \theta(x,y) \ 2K_0\!\left(|k_z|\rho\right)\cos(k_xx+k_yy)\,.
\end{equation}
Here, $K_0(z)$ is the modified Bessel function that diverges as $-\ln(z/2)$ for $z\rightarrow0$ and $\rho=(x^2+y^2)^{1/2}$.  The $xy$ integral is of finite extent so divergences in $\hat{v}^{wi}_c(k)$ originate from $K_0$ as $k_z\to0$.  Using the asymptotic form $K_0(z)= -\ln z + O(z^0)$, we isolate the divergent term
\[
 -2\ln(|k_z|)  \int\! dx\! \int \! dy\, \theta(x,y)\cos(k_xx+k_yy)\,.
\]
For \emph{any} wave vector with $k_{xy}\ne0$, we make this divergence vanish by setting $\theta(x,y)\ne0$ in exactly one Wigner-Seitz cell centered on the wire in the $xy$ plane.  The above $xy$ integral is then zero when $k_{xy}\ne0$ because the projection of $k$ in the $xy$ plane is a reciprocal lattice vector.  This make all entries of $\hat{v}^{wi}_c(k)$ finite when $k\ne0$, and only $k=0$ diverges.  For an arbitrary unit cell in the $xy$ plane, no further simplification is possible and $\hat{v}^{wi}_c(k)$ must be computed numerically.

This choice is superior to what one may have originally used:  cylindrical confinement \cite{CNTvctrunc}
\[
\theta^{cyl}(x,y) = \theta\left(\rho_c - \rho\right)\,.
\]
This cylindrical choice gives
\begin{eqnarray*}
\hat{v}^{cyl}_c(k) & = &  
\frac{4\pi}{k^2}\left\{
1+
k_{xy}\rho_cJ_1(k_{xy}\rho_c)K_0(|k_z|\rho_c)-\right.\\
 & & \left.
\qquad|k_z|\rho_cJ_0(k_{xy}\rho_c)K_1(|k_z|\rho_c)
\right\}
\end{eqnarray*}
Although we have a closed form, for any $k_{xy}\ne0$ the $K_0$ term diverges as $k_z\rightarrow0$.  (This is because a 2D unit cell in the $xy$ plane can not have a cylindrical cross section.)  Many entries $W_{GG'}(q)$ are divergent, and this leads to numerical difficulties.
In practice, we integrate the entries $W_{GG'}(q)$ over $q$, and the integrals are finite. However, performing the integrals requires very dense $q$ sampling \cite{CNTvctrunc}.  In contrast, our mathematically justified choice confines the divergent behavior to the single element $k=0$ in $\hat{v}^{wi}_c(k)$.  As before, this leads to one divergent entry $W_{00}(q)$ that is handled separately.

\begin{figure}[t!]
\includegraphics[angle=-90,width=3.5in]{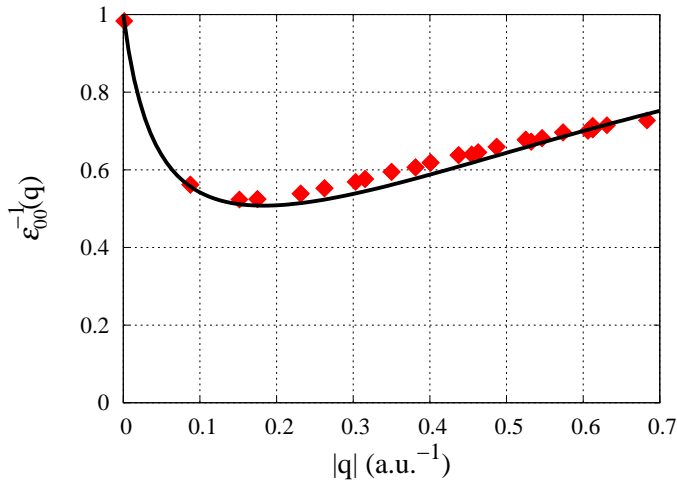}
\caption{\label{fig:fig1}
Static inverse dielectric constant $\epsilon^{-1}_{00}(q)$ versus momentum transfer $q$ for GaN sheets.  The periodic separation between sheets is 14 a.u. and Coulomb truncation was used.  Red diamonds are the first principles results using the RPA.  The solid line is a fit of the form $1/\epsilon^{-1}_{00}(q)=1+\gamma|q|^2\hat{v}^{sh}_c(q)\exp(-\alpha|q|)$ with $\gamma=2.2$ and $\alpha=2.7$ a.u.}
\end{figure}

We illustrate our findings by computing the quasiparticle and optical band gaps of two-dimensional hexagonal sheets of GaN.  These sheets have been investigated with DFT \cite{GaNNTtheo1,GaNNTtheo2,SiandGaNNT,IIINcompareNT}
but not with more accurate excited-state methods. 
In analogy with graphene and carbon nanotubes, these GaN sheets may be the parent materials for single-walled GaN nanotubes, systems of possible interest for their optical and luminescent properties.
The lack of accurate excited-state theory is a serious shortcoming because DFT is known to fail in predicting these very properties \cite{LDAbadgaps,HL}.  

We describe the ground state using DFT-LDA with nonlocal norm-conserving pseudopotentials and a plane-wave cutoff of 35 Ryd \cite{KB,RMP}.  The hexagonal sheets have a primitive lattice vector of 5.92 a.u. in the $xy$ plane.  The supercell is created through k-point sampling of the primitive cell with a regular Monkhorst-Pack grid \cite{MP}.  Within DFT-LDA, the system is insulating with an indirect band gap of 2.4 eV between $\Gamma$ and $K$.  We concentrate on the lowest-energy direct transition which is at $\Gamma$ and has an energy of 3.1 eV within DFT-LDA.  To calculate the excited states, we employ the $GW$ approximation to the self-energy for quasiparticles and the Bethe-Salpeter Equation (BSE) for optical properties \cite{HL,RL}.  The dielectric matrix has a cutoff of 20 Ryd.  Sums over unoccupied states in the $GW$ calculations extends to 22 Ryd.  A static electron-hole interaction is used in the BSE.  All remaining parameters are chosen to converge excitation energies to better than 0.05 eV.

We first consider the quasiparticle band gap at $\Gamma$.  Table \ref{tab:table1}
shows the $GW$ correction to DFT-LDA as a function of k-point sampling and the separation of periodic copies along $z$ when no Coulomb truncation is employed.  The idea is to densely sample in the $xy$ plane to simulate an infinite sheet, consider different periodic separations, and try to find the limit of infinitely-separated sheets.  This corresponds to $n\times n\times1$ k-point sampling of the primitive cell.  Unfortunately, convergence with respect to k-sampling is slow.  Even for very large supercells ($14\times14\times1$), the results are not converged to the typical required accuracy of 0.1 eV.  This difficulty is not related to the k-sampling along $z$: as the Table shows, k-point sampling along $z$ ($n\times n\times m$ sampling) does not improve the situation.  Fundamentally, the long-range image interactions are hindering the convergence.

\begin{table}[t!]
\begin{tabular}{c||c|c|c}
\multicolumn{4}{c}{No truncation:  $GW$ gap correction (eV)}\\
\multicolumn{4}{c}{}\\
k-point & \multicolumn{3}{c}{Sheet separation $L_z$ (a.u.)}\\
sampling & ~~~~10~~~~ & ~~~~14~~~~ & 18\\
\hline
\hline
$4\times4\times1$ & 1.57 & 1.80 & 2.01\\
\hline
$8\times8\times1$ & 1.98 & 2.05 & 2.10\\
\hline
$12	\times12\times1$ & 2.21 & 2.24 & 2.24\\
\hline
$14\times14\times1$ & --- & 2.33 & 2.32\\
\hline
\\
\hline
$8\times8\times1$ & 1.98 & 2.05 & ---\\
\hline
$8\times8\times2$ & 1.78 & 1.99& ---\\
\hline
$8\times8\times3$ & 2.06 & 2.06& ---\\
\hline
\end{tabular}
\caption{\label{tab:table1}
$GW$ correction to the DFT-LDA minimum direct band gap at $\Gamma$ for GaN sheets as a function of k-point sampling and periodic sheet separation.  No Coulomb truncation is employed.  See text for details.}
\end{table}
\begin{table}[t!]
\begin{tabular}{c||c|c|c}
\multicolumn{4}{c}{With truncation:  $GW$ gap correction (eV)}\\
\multicolumn{4}{c}{}\\
k-point & \multicolumn{3}{c}{Sheet separation $L_z$ (a.u.)}\\
sampling & ~~~10~~~ & ~~~14~~~ & 18\\
\hline
\hline
$4\times4\times1$ & 2.39 & 2.60 & 2.83\\
\hline
$8\times8\times1$ & 2.53 & 2.55 & 2.58 \\
\hline
$12\times12\times1$ & 2.53 & 2.53 & 2.52 \\
\hline
$14\times14\times1$ & --- & 2.53 & 2.52\\
\hline
\end{tabular}
\caption{\label{tab:table3}
Same as Table \ref{tab:table1} but with Coulomb truncations.}
\end{table}
\begin{table}[t!]
\begin{tabular}{c||c|c}
\multicolumn{3}{c}{With truncation:  Exciton binding energy (eV)}\\
\multicolumn{3}{c}{}\\
k-point & \multicolumn{2}{c}{Sheet separation $L_z$ (a.u.)}\\
sampling & ~~~~~10~~~~~ & 14\\
\hline
\hline
$4\times4\times1$ & 2.14 & 2.11 \\
\hline
$8\times8\times1$ & 1.46 & 1.40 \\
\hline
$10\times10\times1$ & 1.37 & 1.32\\
\hline
$12\times12\times1$ & 1.33 & 1.30 \\
\hline
$14\times14\times1$ & --- & 1.30 \\
\hline
\end{tabular}
\caption{\label{tab:table4}
Binding energy for the lowest optically active exciton in the GaN sheets.  Coulomb truncation is employed.}
\end{table}
By contrast, our truncation method yields well-converged results with modest effort.  Before presenting the results, we summarize the minor changes required  to treat the diverging element $W_{00}(q)$ in the $GW$ and BSE calculations.  The plan is identical to the case with no truncation:  we replace the divergent element by an appropriate average over $q$. In the $GW$ calculations, we average $W_{00}(q)$ over the first Brillouin zone of the supercell \cite{HL}.  The average can be performed easily by using a model fitted to the \emph{ab initio} results. We have $W_{00}(q)=\epsilon^{-1}_{00}(q)\hat{v}^{sh}_c(q)$ where we have calculated $\epsilon^{-1}_{00}(q)$ on a grid of nonzero $q$ vectors.  In addition, simple analytical considerations show that for insulators $1/\epsilon^{-1}_{00}(q) = 1 + \hat{v}^{sh}_c(q)f(q)$ where $f(q)$ is smooth and has a Taylor series beginning at quadratic order \cite{CPM}.  A simple form $f(q)=\gamma |q|^2 \exp(-\alpha |q|)$ with two positive parameters $\alpha$ and $\gamma$ is sufficient.  See Figure \ref{fig:fig1} for an illustration.  The fit need be good only in the region of small $q$ since the Brillouin zone of the supercell is a small region close to $q=0$.  For the BSE, similar types of averages are needed \cite{RL}, and we also use the model function.

Table \ref{tab:table3} shows that our Coulomb truncation method leads to rapid convergence.  We conclude that the $GW$ correction for an isolated sheet is 2.5 eV, making for a total direct quasiparticle band gap of 5.6 eV at $\Gamma$.  Table \ref{tab:table4} displays the convergence of the binding energy of the lowest-energy exciton calculated with the BSE.  This exciton is optically allowed.  We conclude that the binding energy is 1.3 eV.  Thus the lowest exciton for the sheet is has an excitation energy of 4.3 eV and is optically bright.  However, some caution is required before extrapolating this result to the optical properties of GaN nanotubes.  In semiconducting carbon nanotubes, dark (diople-forbidden) excitons are the lowest in energy likely leading to detrimental reduction of luminescence efficiency \cite{CNTlifetime}.  Our preliminary calculations on GaN nanotubes point to the same problem.

In summary, we derive a scheme for truncating the Coulomb interaction in plane-wave calculations of systems with confined electronic states.  The method excludes periodic image interactions and yields well-converged results for modestly sized periodic cells.  We illustrate the method by using it to calculate the lowest energy optical excitations in two-dimensional hexagonal GaN sheets.

Computer resources were provided by Yale High Performance Computing.  This work was facilitated by the U.S. Department of Energy's Computational Materials Science Network (CMSN).


\begin{thebibliography}{99}
\bibitem{insulcomment} These considerations apply to insulating systems where the divergent ``head'' element presents difficulties while the ``wings'' (where either $G$ or $G'$ but not both are zero) do not contribute as $q\to0$ \cite{HL}.  For metallic systems, the long-range screening ensures that all elements of $W_{GG'}(q)$ remain finite as $q\to0$, i.e. the screened Coulomb interaction is short-ranged in real space.  Thus, for metals, our Coulomb truncation scheme is still useful but may not lead to greatly accelerated convergence.
\end{thebibliography}
\end{document}